\documentclass[aps,pra,reprint,superscriptaddress,nofootinbib,longbibliography]{revtex4-2}

\usepackage{amsmath,amssymb,bm}
\usepackage{physics}
\usepackage{braket}
\usepackage{graphicx}
\usepackage{xcolor} 
\usepackage{color}

\usepackage[ruled,linesnumbered]{algorithm2e}

\usepackage[colorlinks=true,citecolor=blue,linkcolor=blue,urlcolor=blue]{hyperref}

\begin{document}

\title{Adaptive Quantum Metrology of Gravitational Waves}

\author{Wan-Ting He}
\affiliation{School of Optics and Photonics, Beijing Institute of Technology, \\
and Center for Photonic Quantum Precision Measurement (Beijing Institute of Technology), Beijing 100081, China}

\author{Shuang-Quan Ma}
\thanks{\resizebox{\dimexpr\linewidth-1em\relax}{!}{Corresponding authors: These authors jointly directed the project.}\\
mashuangquan@mail.bnu.edu.cn, xukemi@bit.edu.cn}
\affiliation{School of Physics and Astronomy, Beijing Normal University,\\
and Key Laboratory of Multiscale Spin Physics (Beijing Normal University), Ministry of Education, Beijing 100875, China}

 \author{Ke-Mi Xu}
\thanks{\resizebox{\dimexpr\linewidth-1em\relax}{!}{Corresponding authors: These authors jointly directed the project.}\\
mashuangquan@mail.bnu.edu.cn, xukemi@bit.edu.cn}
 \affiliation{School of Optics and Photonics, Beijing Institute of Technology, \\
 and Center for Photonic Quantum Precision Measurement (Beijing Institute of Technology), Beijing 100081, China}


\begin{abstract}
We develop a feedback-controlled quantum metrology scheme to probe weak gravitational waves (GWs) through their direct energy exchange with light. Adaptive SU(2) direction control, based on current estimates of the GW frequency and phase, allows the frequency shifts acquired in successive interaction intervals to add rather than cancel. The resulting persistent frequency difference is converted into an optical phase difference for interferometric readout. We further extend the scheme to elliptically polarized GWs using two transverse sensing channels separated by $45^\circ$ to probe the $+$ and $\times$ components independently. After the GW interaction, the two signals are mapped onto orthogonal polarization modes, enabling phase accumulation along a common path. We derive the corresponding quantum Fisher information matrix and find a $T_{\rm int}^2T_{\rm ph}^2$ scaling in the ideal coherent regime. For a fixed-photon-number probe, the sensitivity shows Heisenberg-like $1/n$ scaling in the ideal case.
\end{abstract}

\maketitle

\section{Introduction}
Laser interferometers detect gravitational waves (GWs) primarily through optical phase changes induced by spacetime strain~\cite{Abbott2016,Aasi2015,Acernese2015,Bailes2021}. 
Their sensitivity depends on the effective optical path length and the suppression of classical and quantum noise~\cite{Danilishin2012,Freise2016,Danilishin2019,Adhikari2014}. 
Quantum fluctuations of light limit the sensitivity of advanced interferometric detectors, motivating the development of quantum measurement techniques~\cite{Caves1981,Kimble2002,Schnabel2010,Danilishin2019}.
Quantum optical techniques have become an important part of this development. In particular, squeezed states of light have been used to reduce quantum noise in LIGO and Virgo, and frequency-dependent squeezing
now provides broadband quantum enhancement~\cite{Abadie2011,Aasi2013,Dwyer2013,Dooley2015,Tse2019,Acernese2019,McCuller2023}.

Alternative interferometer topologies, including Sagnac and speed-meter configurations, have also been studied for broadband and quantum-noise-limited GW detection~\cite{BraginskyKhalili1990,Braginsky2000,Sun1996,Hogan2001,
Chen2003,Danilishin2004,Eberle2010,Graef2014,Danilishin2015,
Huttner2017,Danilishin2018}. More generally, quantum-noise engineering, entanglement-assisted readout, and modified interferometer dynamics have been explored as possible routes beyond the standard quantum limit~\cite{Miao2014,Miao2017,Ma2017,KhaliliPolzik2018,Danilishin2019}.

GW-induced photon-frequency shifts and quantum GW--optical interactions have been studied in several settings~\cite{Kaufmann1970,MendoncaDrury2001,PangChen2018,Bringmann2023,Lobato2021,TintoArmstrong1998,Kolkowitz2016,Guerreiro2020,Arani2025}. Magnetically assisted graviton--photon conversion has also been explored for electromagnetic GW sensing~\cite{DomckeGarciaCely2021,DomckeGarciaCelyRodd2022,Berlin2022,Ikeda2025,DeLogiMickelson1977,Herman2021,Herman2023,ItoKohriNakayama2024,LiuRenZhang2024}. A complementary approach probes the direct exchange of energy between a pre-existing GW and light~\cite{Schuetzhold2025}. Sch\"utzhold showed that a Mach--Zehnder- or Sagnac-type geometry can convert the oscillating GW-induced optical-frequency modulation into a persistent frequency shift by changing the propagation direction of the light at selected phases of the GW cycle~\cite{Schuetzhold2025}. The switching sequence determines whether the optical field gains or loses energy. The more general question of energy exchange between gravitational
fields and quantum systems, as well as the possibility of detecting
individual gravitational quanta, has been considered in several
different settings~\cite{BoughnRothman2006,RothmanBoughn2006,Giovannini2019,
Graefe2023,Spengler2022,Tobar2024,Tobar2025}. 
If the weak gravitational field is quantized, the energy exchange between the GW and the optical field can be interpreted as stimulated graviton absorption or emission~\cite{Schuetzhold2025}. In the present work, we focus on the optical response itself. 

Implementing this response requires switching synchronized with the GW frequency and phase. Errors in the estimated frequency and phase therefore reduce the accumulated response~\cite{Schuetzhold2025}. In general relativity, a GW has two tensor polarization components, $h_+(t)$ and $h_\times(t)$~\cite{FlanaganHughes2005}. In the geometry considered here, a single optical channel measures only one direction-dependent combination of the two components. 
Quantum parameter estimation provides a natural framework for these problems. The quantum Fisher information (QFI) sets precision bounds through the probe’s dependence on the unknown parameters~\cite{Helstrom1976,Holevo1982,BraunsteinCaves1994,
Giovannetti2004,Giovannetti2011,Pezze2018,YuanFung2017}.

For general dynamics, including time-dependent evolution, the  generators of parameter translations can change direction over time. As a result, information gained at different times may fail to accumulate coherently~\cite{PangBrun2014,Liu2015,Jing2015,YuanFung2017,PangJordan2017}.
Similar issues arise in waveform estimation and continuous sensing of time-varying signals~\cite{TsangWisemanCaves2011,Tsang2012}.
Quantum control can modify this evolution to preserve
the accumulation of parameter information. Quantum control and probe--environment engineering can preserve parameter information and improve weak-signal sensitivity~\cite{Long2022Zeno,Dong2025Zeno,Dong2026Axion}. Feedback and sequential control have been studied for Hamiltonian parameter estimation~\cite{YuanFung2015,Yuan2016,PangJordan2017,Demkowicz2017,Titum2021,Yang2024}, and control-enhanced sequential metrology has been experimentally demonstrated near the Heisenberg limit~\cite{Hou2019}. Sequential phase-estimation experiments have also approached Heisenberg-like precision without relying exclusively on large multiparticle entangled states~\cite{Higgins2007,Xiang2011,Slussarenko2017,Daryanoosh2018}.

These ideas motivate a feedback-controlled extension of the GW--light energy-exchange scheme. We first consider a single projected GW amplitude. Light propagating along two orthogonal directions
acquires opposite frequency shifts. The GW frequency $\omega$ and phase $\phi$ are estimated from earlier probe measurements, and the updated values are used to determine the switching times for the following measurement. SU(2) control then exchanges the two propagation directions at these switching times. When the switching is well synchronized with the GW, the frequency shifts acquired during successive parts of the GW cycle add rather than cancel. As in adaptive sequential quantum metrology, the control preserves coherent accumulation of the parameter-dependent response~\cite{YuanFung2015,Yuan2016,PangJordan2017,Hou2019}. The control does not increase the instantaneous GW–light coupling.

We then extend the scheme to both GW polarization amplitudes, whose separation requires an additional response channel. For normal incidence, two channels rotated by $45^\circ$ give orthogonal responses to the + and $\times$ components. In this geometry, the two-polarization problem takes a particularly simple form and can be treated as two parallel single-parameter sensing channels. This setting also provides a natural route to quantum multiparameter estimation. We derive the QFI for a single GW amplitude and then generalize the result to the two-parameter QFI matrix (QFIM).

Under synchronized feedback control, the frequency response grows with the interaction time $T_{\rm int}$, while the persistent frequency difference is converted into phase over the delay time $T_{\rm ph}$. The resulting QFI therefore scales as $T_{\rm int}^2T_{\rm ph}^2$ in the ideal coherent regime. In the $45^\circ$ geometry, the two amplitudes are encoded in separate optical channels, which leads to compatible local estimation. For a chosen probe, the ideal  sensitivity further shows Heisenberg-like $1/n$ scaling with the probe photon number $n$ for large $n$.

This paper is organized as follows. In Section~\ref{section II}, we introduce the single-channel GW--light interaction and its optical-frequency response. In Section~\ref{section III}, we develop the feedback-controlled sensing scheme and analyze the single-parameter quantum sensitivity. In Section~\ref{section 4}, we extend the scheme to elliptically polarized GWs and derive the two-parameter QFIM. In Section~\ref{section 5}, we summarize the main results.

\section{Single-Channel GW--Light Interaction}\label{section II}

We begin with the single-polarization geometry considered in Ref.~\cite{Schuetzhold2025}. In the transverse-traceless gauge, a linearly polarized GW propagating along the $z$ direction is described by the metric
\begin{equation}
ds^2=dt^2-[1+h(t)]dx^2-[1-h(t)]dy^2-dz^2 .
\end{equation}
We assume a monochromatic GW strain of the form 
\begin{equation}\label{eq:GW_strain}
h(t)=A\cos(\omega t-\phi), 
\end{equation}
where $A$, $\omega$, and $\phi$ denote its amplitude, angular frequency, and phase, respectively.

We then describe an optical field polarized along the $z$ direction by the vector $\bm A(t,\bm r)=A_z(t,x,y)\bm e_z$.
The optical frequency $\Omega$ is assumed to be much larger than the GW frequency, i.e., $\Omega\gg\omega$. The GW background therefore changes negligibly during a single optical cycle, allowing the electromagnetic field to be treated within the Wentzel--Kramers--Brillouin (WKB) approximation~\cite{Schuetzhold2025}. Between reflections, the transverse wave vector is approximately conserved, while the optical frequency acquires a slow time dependence induced by the GW background. 

The GW-induced frequency shift depends on the propagation direction of the light in the transverse plane. To describe this directional dependence, we write the transverse wave vector as $\bm K=K(\cos\gamma,\sin\gamma,0)$, where $\gamma$ is the propagation angle in the x-y plane. The corresponding  WKB dispersion relation is $\Omega_\gamma^2(t)=K^2[1-h(t)\cos 2\gamma]$. For a weak GW with $|h|\ll1$, the dispersion relation reduces to 
\begin{equation}\label{eq:dispersion}
\Omega_\gamma(t)\simeq K\left[1-\frac{h(t)}{2}\cos2\gamma\right].
\end{equation}
For propagation along the $x$ and $y$ directions, i.e.,
$\gamma=0$ and $\gamma=\pi/2$, respectively, the corresponding
dispersion relations are given by
\begin{subequations}
\label{eq:Omega_xy}
\begin{align}
\Omega_x(t) &\simeq K\left[1-\frac{h(t)}{2}\right], \label{eq:Omega_x}\\
\Omega_y(t) &\simeq K\left[1+\frac{h(t)}{2}\right]. \label{eq:Omega_y}
\end{align}
\end{subequations}
Thus, the GW produces equal and opposite frequency shifts along the two orthogonal directions. These frequency shifts correspond to energy exchange between the GW and the optical field.

During free propagation, the photon number is conserved, and the optical energy is $E=N\hbar\Omega$. Equations~\eqref{eq:Omega_x} and~\eqref{eq:Omega_y} therefore give
\begin{subequations}
\begin{align}
\frac{\dot E_x}{E_x}=&\frac{\dot\Omega_x}{\Omega_x}
\simeq-\frac{1}{2}\dot h(t),\\
\frac{\dot E_y}{E_y}=&\frac{\dot\Omega_y}{\Omega_y}\simeq+\frac{1}{2}\dot h(t).
\end{align}
\label{eq:single_energy_rate}
\end{subequations}
The two propagation directions yield opposite rate of energy changes, consistent with the GW--light energy-transfer relation derived in Ref.~\cite{Schuetzhold2025}.

For a fixed propagation direction, $\dot h(t)$ changes sign during each GW cycle. The frequency shifts from different parts of the cycle
therefore tend to cancel. This cancellation can be reduced by switching the propagation direction when the GW response changes sign. Since the
$x$ and $y$ directions have opposite responses, the frequency shifts can then accumulate with the same sign.

To obtain a differential signal, we consider two optical branches with complementary propagation paths. One branch propagates along $x$ and the other along $y$,  with their directions switched simultaneously. The two branches then acquire opposite frequency shifts, yielding the frequency difference
\begin{equation}\label{eq:Delta_Omega}
\Delta\Omega=\Omega\mathcal{R}A,
\end{equation}
where $\mathcal{R}$ is a dimensionless response factor determined by the GW waveform and the switching sequence.

\section{Control-Enhanced Single-Parameter Estimation}\label{section III}

The switching sequence depends on the GW frequency and phase. Uncertainty in these parameters can cause timing errors that reduce the accumulated frequency shift. We therefore use feedback control to update the switching times during the measurement.
The GW frequency and phase are estimated from earlier probe pulses. The updated values, $\omega$ and $\phi$, are then used to set the switching times for the next measurement.

The two propagation directions are treated as two optical modes with annihilation operators $\hat a_x$ and $\hat a_y$. In the Schwinger representation, we define
\begin{subequations}
\begin{align}
\hat J_2=&\frac{1}{2i}\left(\hat a_x^\dagger\hat a_y-\hat a_y^\dagger\hat a_x\right),\\
\hat J_3=&\frac{1}{2}\left(\hat a_x^\dagger\hat a_x-\hat a_y^\dagger\hat a_y\right).
\end{align}
\end{subequations}
Here, $\hat J_2$ generates the exchange between the two propagation directions, and $\hat J_3$ describes their differential response to the GW. 

According to Eqs.~\eqref{eq:Omega_x} and~\eqref{eq:Omega_y} , the optical Hamiltonian can be written as
\begin{equation}\label{Hamiltonian}
\hat H(t) =\hbar\Omega\hat N-\hbar\Omega h(t)\hat J_3,
\end{equation}
where $\hat N=\hat a_x^\dagger\hat a_x+\hat a_y^\dagger\hat a_y$. 
The first term of Eq.~\eqref{Hamiltonian} generates a common optical evolution and does not contribute to the differential signal. The GW interaction Hamiltonian is therefore 
$\hat H_{\rm GW}(t)=-\hbar\Omega h(t)\hat J_3 $.

To describe the feedback process, we divide the interaction time $T_{\rm int}$ into $N$ intervals of duration, $\tau=T_{\rm int}/N$.
During the $k$th interval, the GW-induced evolution is
\begin{equation}
\hat U_{\rm GW}^{(k)}
=
\exp
\left(
i\chi_k\hat J_3
\right),
\label{eq:U_GW}
\end{equation}
where $\chi_k=(\Omega A/\omega)[\sin(\omega(t_k+\tau)-\phi)-\sin(\omega t_k-\phi)]$.
Each interaction interval carries information about the GW amplitude, frequency, and phase.

Between two interaction intervals, we apply an SU(2) control $\hat U_c^{(k)}=e^{-i\theta_k\hat J_2}$, where $\theta_k$ is the control angle. For a time-dependent control,
the corresponding Hamiltonian is $\hat H_c(t) = \hbar\dot{\theta}(t)\hat J_2$.
A $\pi$ rotation exchanges the $x$ and $y$ propagation directions.
This rotation reverses the sign of $\hat J_3$, satisfying $\hat U_c^\dagger(\pi)\hat J_3\hat U_c(\pi)=-\hat J_3$. The control therefore reverses the sign of the optical GW response.

The full interaction can be described by the sequence of GW evolution and control operations
\begin{equation} 
\hat U_{\rm int} = \hat U_{\rm GW}^{(N)} \hat U_c^{(N-1)} \cdots \hat U_c^{(2)} \hat U_{\rm GW}^{(2)} \hat U_c^{(1)} \hat U_{\rm GW}^{(1)} .
\end{equation}
This sequence follows the sequential-feedback approach to quantum parameter estimation~\cite{Yuan2016,Hou2021}. Here, feedback acts on the two optical propagation modes.

Let $t_0,t_1,\ldots,t_N$ denote consecutive switching times. For complementary optical branches, the dimensionless frequency response follows the simple recurrence
\begin{equation}
\mathcal R_{k+1}\!=\!\mathcal R_k\!+\!(-1)^k
\big[\cos(\omega t_k-\phi)\!-\!\cos(\omega t_{k+1}-\phi)\big],
\end{equation}
with $\mathcal R_0=0$. 
For ideal switching at $t_k=(\phi+k\pi)/\omega$, each interval increases the response by $2$. The accumulated response and frequency difference are therefore
\begin{subequations}
\begin{align}
\mathcal R_N=&2N=\frac{2\omega T_{\rm int}}{\pi},\\
\Delta\Omega=&\Omega A\mathcal R_N,
\end{align}
\end{subequations}
where $T_{\rm int}=N\pi/\omega$ and $N$ is an integer. The control enhances the accumulated response by reducing cancellation between frequency shifts, without increasing the instantaneous GW--light coupling.



After the GW--light interaction, the residual frequency difference accumulates an additional relative phase during $T_{\rm ph}$. The total phase $\Phi(A)$ is 
\begin{equation}
\Phi(A)=\Omega A\left[ \sum_{k=0}^{N-1}\int_{t_k}^{t_{k+1}}\mathcal R(t)\,dt+\mathcal R_NT_{\rm ph}\right].
\end{equation}
For ideal synchronized switching, $\mathcal R_k=2k$. The interaction integral equals $\sum_{k=0}^{N-1}(2k+1)\pi/\omega=\pi N^2/\omega$, which gives $\Phi(A)=\Omega A\omega \left(T_{\rm int}^2+2T_{\rm int}T_{\rm ph}\right)/\pi$. In the long-delay regime, $T_{\rm ph}\gg T_{\rm int}$, the phase accumulated after the interaction dominates, giving $\Phi(A)\propto \Omega\omega T_{\rm int}T_{\rm ph}A$.

\begin{figure}[t]
    \centering    \includegraphics[width=1\linewidth, clip=true, trim=10 120 5 100]{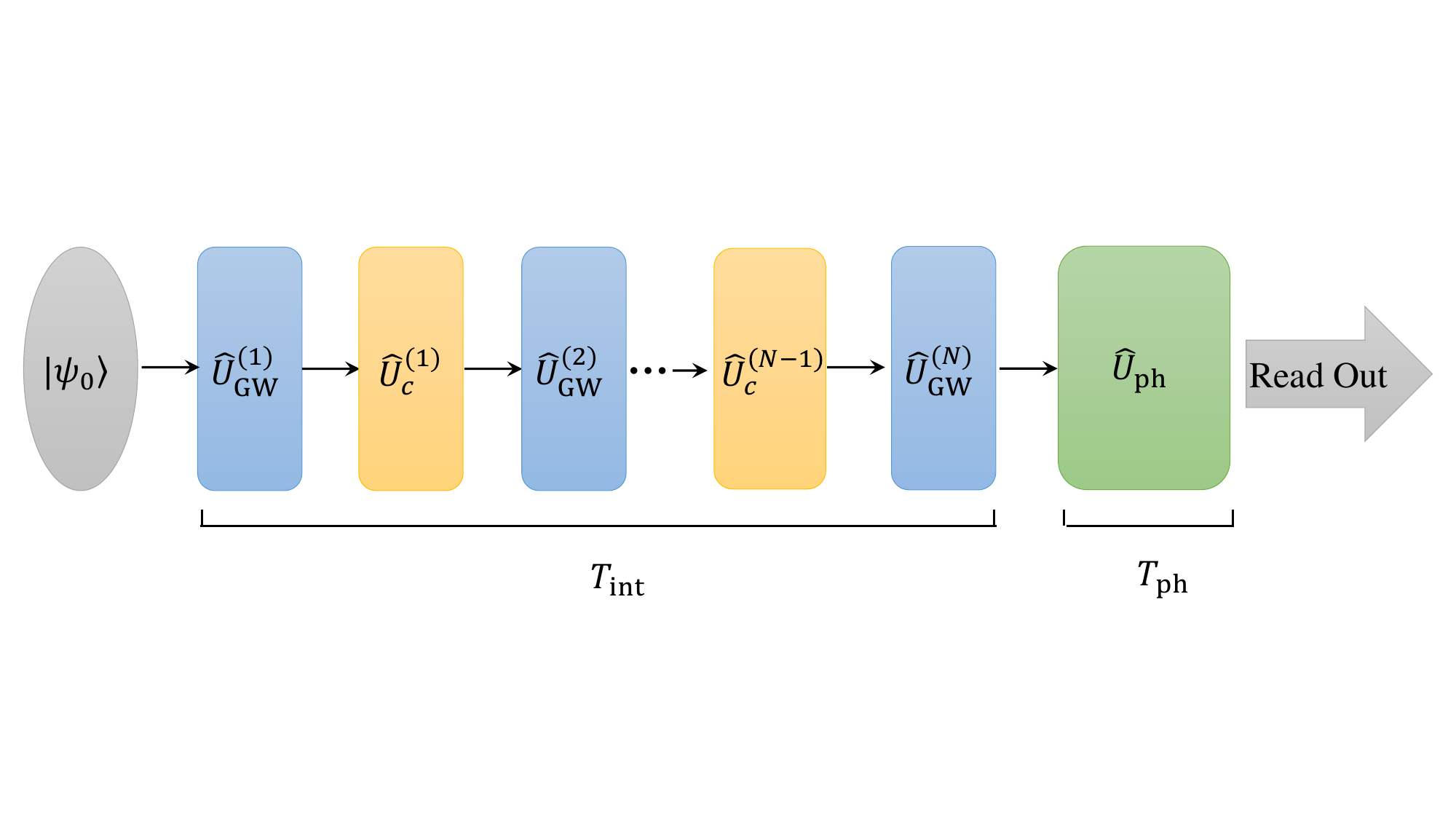}
    \caption{
    Schematic of the feedback-controlled GW sensing protocol.   During the interaction time $T_{\rm int}$, the probe undergoes alternating GW-induced evolutions $\hat U_{\rm GW}^{(k)}$ and
    control operations $\hat U_c^{(k)}$, where the switching times are determined from the current estimates of the GW frequency and  phase. After the last GW-interaction interval, the accumulated
    frequency difference is converted into a relative phase by  $\hat U_{\rm ph}$ during the coherent phase-accumulation time  $T_{\rm ph}$, followed by the final readout.
    }
    \label{fig1}
\end{figure}

Let $\hat G$ denote the relative-phase generator of the readout modes. The phase accumulation is described by $\hat U_{\rm ph} = e^{-i\Phi(A)\hat G}$.
The complete feedback-controlled sensing protocol is summarized in Fig.~\ref{fig1}.
 
The two stages play different roles in the sensing process. The sequence within $T_{\rm int}$ converts the oscillating GW response into an accumulated frequency difference, while $\hat U_{\rm ph}$
converts this frequency difference into a measurable phase during $T_{\rm ph}$. 

We now quantify the precision of GW amplitude estimation using this
phase encoding introduced above. For the two readout modes $\hat b_1$ and $\hat b_2$, the relative-phase generator is $\hat{G}=(\hat b_1^\dagger\hat b_1 -\hat b_2^\dagger\hat b_2)/2$. For an $A$-independent pure probe state and fixed feedback settings, the QFI is
\begin{equation}\label{eq:QFI}
 F_A = 4\left(\frac{\partial\Phi}{\partial A}\right)^2\mathrm{Var}(\hat G) \propto \Omega^2\omega^2T_{\rm int}^2T_{\rm ph}^2\mathrm{Var}(\hat G).
\end{equation}
Here $T_{\rm int}^2$ arises from the accumulated GW response, whereas $T_{\rm ph}^2$ arises from the subsequent phase accumulation. The probe state determines the remaining factor $\mathrm{Var}(\hat G)$.

\begin{figure}[t]
    \centering
    \includegraphics[width=1\linewidth, clip=true, trim=20 40 10 5]{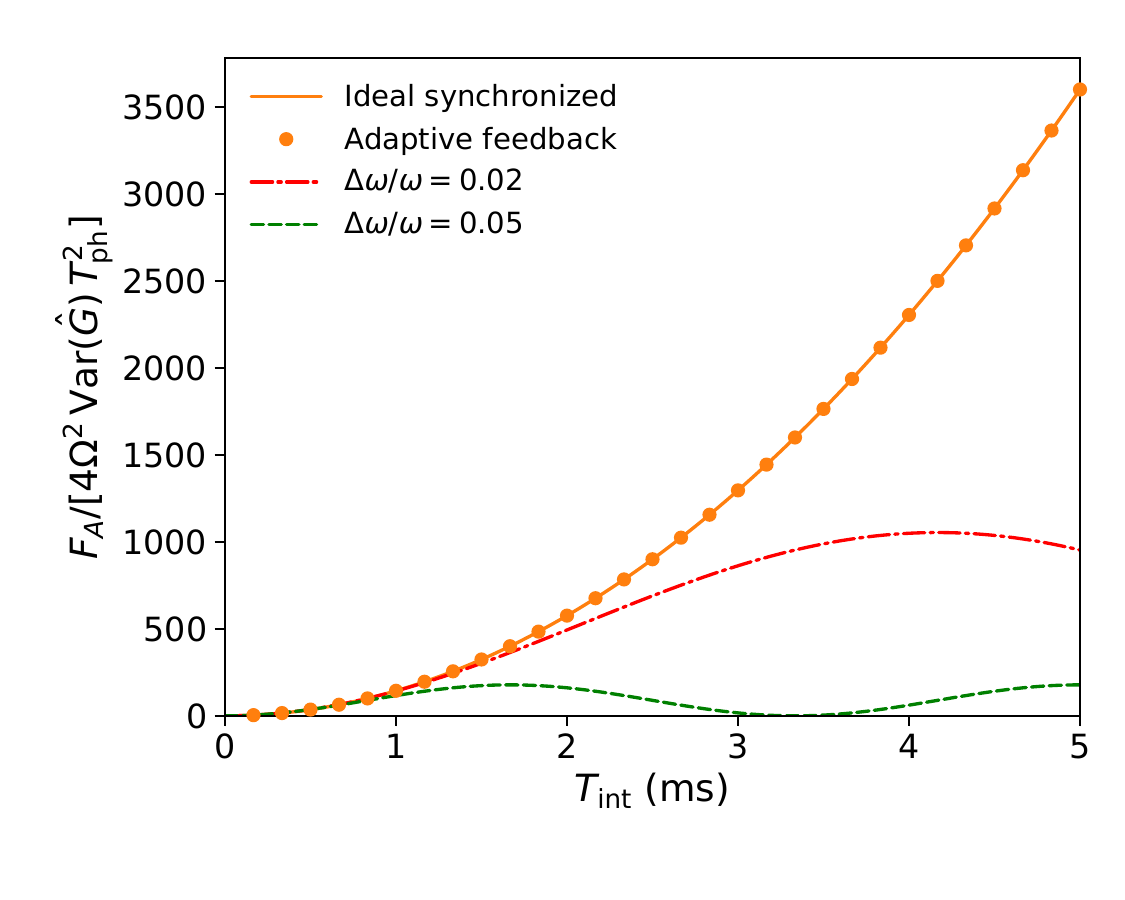}
\caption{The normalized QFI
$F_A/[4\Omega^2\mathrm{Var}(\hat G)T_{\rm ph}^2]$
is plotted against $T_{\rm int}$ for
$\omega/(2\pi)=3\,\mathrm{kHz}$.
The orange solid line shows the ideal synchronized result, while
orange circles indicate numerical values at
$T_{\rm int}=N\pi/\omega$, representing the synchronized limit of
adaptive feedback.
The red dash-dotted and green dashed curves show smooth trends
through the switching-time responses for preset control with
$\Delta\omega/\omega=0.02$ and $0.05$, respectively,
where $\Delta\omega$ denotes the preset control-frequency error.
Synchronized control preserves quadratic growth with interaction
time, whereas frequency mismatch produces oscillatory cancellation.
}
\label{fig:qfi_time}
    \label{fig2}
\end{figure}

To isolate the contribution of the feedback-controlled GW response, we use the normalized QFI $F_A/[4\Omega^2\mathrm{Var}(\hat G)T_{\rm ph}^2]$. The numerical results are shown in Fig.~\ref{fig2}. For preset switching with $\Delta\omega/\omega=0.02$ and $0.05$, the normalized QFI is strongly reduced from the ideal synchronized result. In contrast, adaptive feedback closely follows the ideal case over the whole interaction time.

With synchronized feedback, the switching sequence stays matched to the GW waveform, so the responses from successive interaction intervals add coherently. The accumulated response therefore grows approximately linearly with $T_{\rm int}$, leading to the quadratic scaling $\mathcal F_A\propto T_{\rm int}^2$.
For preset switching, the frequency shift causes the switching times to gradually drift away from the optimal GW phases. The responses from different intervals then partially cancel, producing the oscillations in the normalized QFI.

We next consider the contribution from the probe state. For a fixed total photon number $n=2j$, we choose the zero-eigenvalue state $|j,0\rangle_2$ of $\hat J_2$ as the probe state, i.e.,  $|\psi_0\rangle=|j,0\rangle_2$. If the readout mapping identifies $\hat{G}$ with $\hat{J}_3$, the phase-generator variance is
\begin{equation}\label{eq:Var}
\mathrm{Var}(\hat G)=\frac{j(j+1)}{2}=\frac{n(n+2)}{8},
\end{equation}
which grows as $n^2$ for large photon number. Substituting Eq.~\eqref{eq:Var} into Eq.~\eqref{eq:QFI} yields 
\begin{equation}
F_A \propto \Omega^2\omega^2 T_{\rm int}^2T_{\rm ph}^2 n(n+2).
\end{equation}
For $\nu$ independent measurements, the quantum Cram\'er--Rao bound is
\begin{equation}
\Delta^2 A \geq \frac{1}{\nu F_A}.
\label{eq:17}
\end{equation}
Using $n(n+2)\simeq n^2$ for $n\gg1$, the QFI and the uncertainty in the GW
amplitude respectively scale as 
\begin{subequations}
\begin{align}
F_A \propto &\Omega^2\omega^2 T_{\rm int}^2T_{\rm ph}^2 n^2,\\
\Delta A\propto&\frac{1}{\sqrt{\nu}\,\Omega\omega T_{\rm int}T_{\rm ph}n}.
\end{align}
\end{subequations}
\begin{figure}[hbt!]
    \centering
    \includegraphics[width=1\linewidth, clip=true, trim=0 0 0 0]{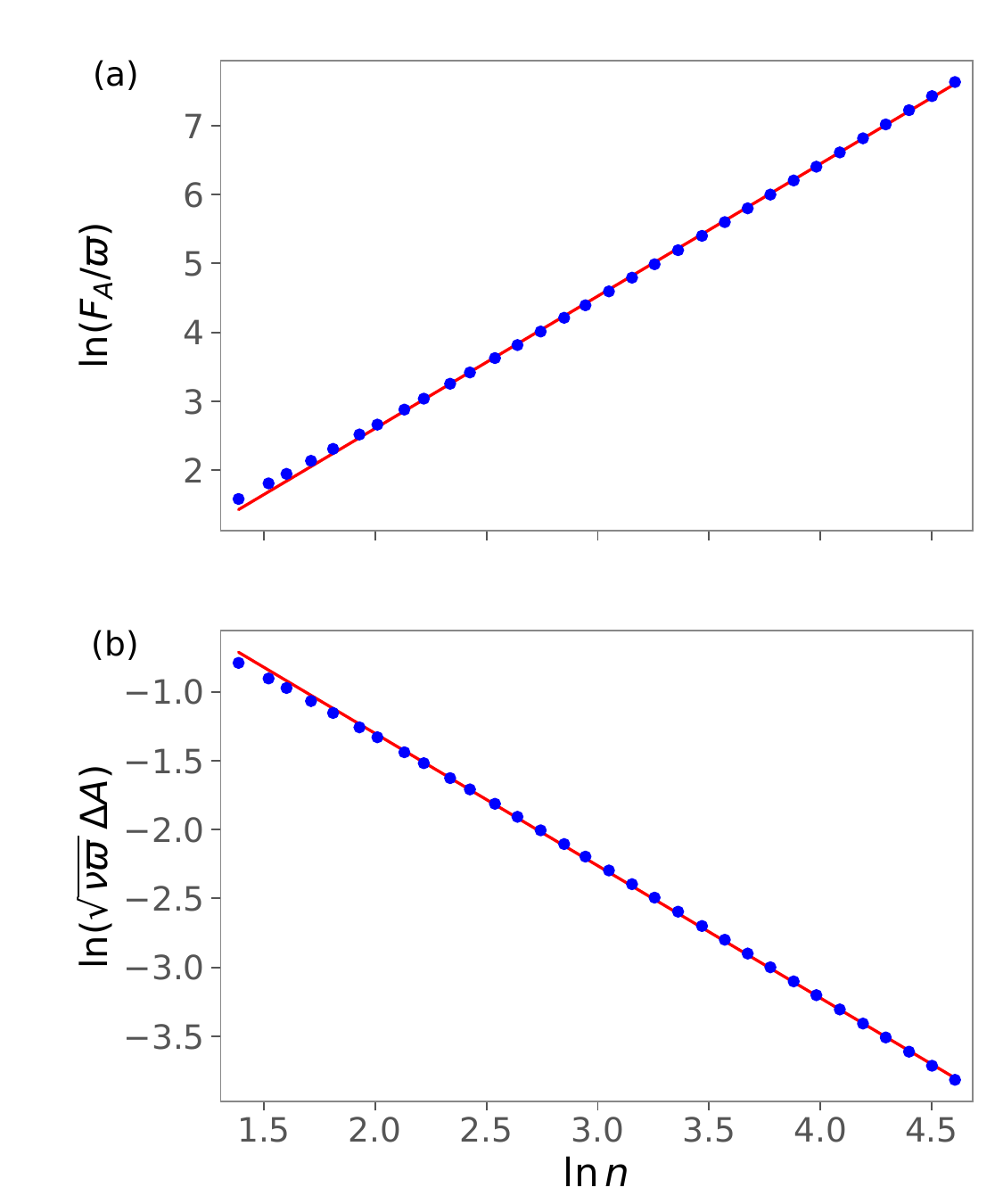}
\caption{
Photon-number dependence of the QFI and
the uncertainty for the probe state
$|\psi_0\rangle=|j,0\rangle_2$, with total photon number $n=2j$,
under ideal synchronized control in the limit
$T_{\rm ph}\gg T_{\rm int}$.
(a) Normalized QFI
$F_A/\varpi$, where $\varpi=\Omega^2\omega^2T_{\rm int}^2T_{\rm ph}^2$.
(b) The corresponding amplitude-estimation uncertainty $\Delta A$ obtained from the same QFI using Eq.~\eqref{eq:17}. Both panels use natural-logarithmic coordinates. Blue circles show representative theoretical values, and red solid lines show linear least-squares fits in logarithmic coordinates over $4\leq n\leq100$.
The fitted slopes are approximately $1.920$ and $-0.960$,
respectively, with the latter equal to minus one-half of the former.
For $n\gg1$, these relations approach $F_A\propto n^2$ and
$\Delta A\propto n^{-1}$ at fixed measurement number and
sensing parameters.
}
\label{figre31}
\end{figure}
Thus, the chosen probe gives the Heisenberg-like $n^2$ and $1/n$ scalings in the ideal case, while a coherent probe with $\mathrm{Var}(\hat G)=O(n)$ gives the usual $1/\sqrt n$ scaling. 
Figure~\ref{figre31}(a) shows the photon-number dependence of the
QFI for the chosen probe under ideal synchronized control.
The linear fit in logarithmic coordinates over
$4\leq n\leq100$ has a slope of approximately $1.920$.
This finite-range slope follows from the dependence
$F_A\propto n(n+2)$, which approaches quadratic scaling, i.e., $F_A\propto n^2$, for
$n\gg1$. Figure~\ref{figre31}(b) shows the corresponding uncertainty. The logarithmic fit over the same photon-number range has a slope of approximately $-0.960$, as required by the quantum
Cram\'er--Rao relation. For $n\gg1$, the uncertainty bound approaches
$\Delta A\propto1/n$, yielding Heisenberg-like photon-number
scaling of the ideal amplitude sensitivity.

\section{Two-Polarization Estimation}\label{section 4}

We now extend the frequency-transport protocol to the simultaneous estimation of two GW polarization amplitudes. 
For a monochromatic elliptically polarized GW propagating along the $z$ direction, the two polarizations are 
\begin{subequations}\label{eq:projection}
\begin{align}
h_+(t) =& A_+\cos(\omega t-\phi),\\
h_\times(t) =& A_\times\sin(\omega t-\phi),
\end{align}
\end{subequations}
where the amplitudes $A_+$ and $A_\times$ form the parameter vector $\bm A= (A_+,A_\times)^{\mathsf T}$ to be estimated. 
A single nonzero component gives linear polarization, whereas $|A_+|=|A_\times|\ne0$ gives circular polarization. Hereafter, $\Theta\in\{+,\times\}$ labels the polarization channel, and relations indexed by $\Theta$ apply to each channel separately. A sensing channel oriented at an angle $\gamma$ in the transverse plane responds to the projected strain
\begin{equation}
 h_\gamma(t)=h_+(t)\cos(2\gamma)+h_\times(t)\sin(2\gamma).
\end{equation}
The channel orientation determines the relative contributions of the two polarization components. $\gamma=0$ selects the $+$ component, whereas $\gamma=\pi/4$ selects the $\times$ component. 

We consider two sensing channels with independently controlled switching sequences. 
The first channel, $(\gamma_1,\chi_1)=(0,0)$, serves as a fixed reference, while the second channel has orientation $\gamma_2=\theta$ and relative switching delay $\delta t$, with $\chi=\omega\,\delta t$. 
We take $\omega$ and $\phi$ as calibrated and let both channels interrogate $N$ complete GW half-cycles. The two probes are independent pure states whose relative-phase generators have the same variance $V$. We work in the long-delay regime, $T_{\rm ph}\gg T_{\rm int}$, with fixed interrogation durations and optical resources. The optimization therefore varies the relative orientation $\theta$ and switching delay $\delta t$.


We introduce the GW phase coordinate $x=\omega t-\phi$ and choose the interaction window $[t_0,t_0+T_{\rm int}]$, where $t_0=\phi/\omega$. With $T_{\rm int}=N\pi/\omega$, this gives $x\in[0,N\pi]$. The relative delay $\chi=\omega\delta t$ enters through the switching function
\begin{equation}
 q_\chi(x)=\rm sgn[\sin(x-\chi)],
\end{equation}
whose sign changes occur at $x=\chi+k\pi$ ($k\in\mathbb Z$), so that $\chi$ directly quantifies the relative switching delay. For the two complementary propagation branches, the normalized differential optical-frequency response reads $\Delta\Omega_\gamma/\Omega=\int_0^{T_{\rm int}} q_\chi(\omega t-\phi)[-\dot h_\gamma(t)]\,dt$.
Substituting Eq.~\eqref{eq:projection} and changing variables from $t$ to $x$ gives
\begin{align}
 \frac{\Delta\Omega_\gamma}{\Omega}
 =& A_+\cos(2\gamma)\int_0^{N\pi}q_\chi(x)\sin x\,dx \nonumber\\
 &-A_\times\sin(2\gamma)\int_0^{N\pi}q_\chi(x)\cos x\,dx .
\end{align}
For $0\le\chi\le\pi$, $q_\chi=-1$ for $x\in (0,\chi)$ and $+1$ for $x \in (\chi,\pi)$ within one half-cycle, which gives $\int_0^\pi q_\chi\sin x\,dx=2\cos\chi$ and $-\int_0^\pi q_\chi\cos x\,dx=2\sin\chi$. Since $q_\chi(x+\pi)=-q_\chi(x)$, both integrands are $\pi$ periodic. Summing over $N$ half-cycles yields 
\begin{equation}\label{eq:general-shift}
\frac{\Delta\Omega_\gamma}{\Omega}  =2N\left[A_+\cos(2\gamma)\cos\chi  +A_\times\sin(2\gamma)\sin\chi\right].
\end{equation}

We now relate the accumulated frequency shifts to the phase responses of the two sensing channels. Following the effective readout model of Sec.~\ref{section III}, we assume that the  accumulated frequency difference is preserved during the subsequent coherent interrogation. In the long-delay limit, post-interaction phase accumulation dominates, so that $\Phi_\gamma\simeq \Delta\Omega_\gamma T_{\rm{int}}$. Using the frequency response of Eq.~\eqref{eq:general-shift}, we define the common response scale
\begin{equation}
 g_0=2N\Omega T_{\rm{ph}}=\frac{2}{\pi}\,\Omega\omega T_{\rm{int}}T_{\rm{ph}} .
\label{eq:g0}
\end{equation}
The first channel gives $\Phi_1=g_0A_+$, while  Eq.~\eqref{eq:general-shift} determines the phase $\Phi_2$ of the second channel. The measured phases $\bm\Phi=(\Phi_1,\Phi_2)^{\mathsf T}$ therefore satisfy $\bm\Phi=\bm S\bm A$, with the response matrix
\begin{equation}
\bm S=g_0\begin{pmatrix}1&0\\a&b\end{pmatrix}.
\label{eq:phase-response-matrix}
\end{equation}
Here, $a=\cos(2\theta)\cos\chi$ quantifies the contribution of $A_+$ to the second channel. The coefficient $b=\sin(2\theta)\sin\chi$ sets the response of that channel to $A_\times$.

Let $\hat G_\Theta$ denotes the phase generator of polarization readout mode $\Theta$, with $\hat G_\Theta=[\hat b_{\Theta,1}^\dagger\hat b_{\Theta,1}-\hat b_{\Theta,2}^\dagger\hat b_{\Theta,2}]/2$. The phase evolution is generated independently in the two optical readout channels,
\begin{equation}
\hat U_{\rm ph} = \exp \left[ -i \left( \Phi_+\hat G_+ + \Phi_\times\hat G_\times \right) \right].
\end{equation}
The two signals are encoded in orthogonal polarization modes, and their phase generators commute, i.e., $[\hat G_+,\hat G_\times]=0$.
For a parameter-independent pure probe and a fixed control record, the QFIM associated with this phase model is
\begin{equation}
 \mathcal F_{\mu\nu}  =4\sum_{r,s}S_{r\mu}S_{s\nu}
 \mathrm{Cov}_{\rm sym}(\hat G_r,\hat G_s),
\end{equation}
where $\mu,\nu\in\{+,\times\}$ and $\mathrm{Cov}_{\rm sym}(X,Y)=\langle XY+YX\rangle/2-\langle X\rangle\langle Y\rangle$. For the independent probes with equal generator variance assumed above, $\mathrm{Cov}_{\rm sym}(\hat G_r,\hat G_s)=V\delta_{rs}$. The QFIM therefore reduces to
\begin{equation}
\bm{\mathcal{F}}=4V\bm{S}^{\mathrm T}\bm{S}.
\end{equation}
Assume that $\mathcal{F}_0=4V g_0^2 =16\Omega^2\omega^2 T_{\rm int}^2 T_{\rm ph}^2 V/\pi^2$,
we obtain 
\begin{equation}                                
\bm{\mathcal F}(\theta,\delta t) =\mathcal F_0\
\begin{pmatrix}                   
 1+a^2&ab\\ab&b^2                               
\end{pmatrix}.                                                
\end{equation}  
The least-sensitive amplitude direction is governed by the smallest eigenvalue, 
\begin{equation}                                                             
\frac{\lambda_{\min}(\bm{\mathcal F})}{\mathcal F_0} \!=\!\frac{1+a^2+b^2\!-\!\sqrt{(1+a^2-b^2)^2+4a^2b^2}}{2}.                              
\end{equation}           

Because the two amplitudes must be estimated simultaneously, optimizing the sensitivity to one of them alone may leave the other poorly resolved. We therefore maximize the Fisher information along the least-sensitive direction in the two-dimensional amplitude space.

For $M$ independent repetitions of the protocol, any unbiased estimator $\bm A$ obeys the matrix quantum Cram\'er--Rao bound
\begin{equation}
    \mathrm{Cov}(\bm A) \succeq \frac1M\bm{\mathcal F}^{-1},
\end{equation}
provided that the QFIM $\bm{\mathcal F}$ is nonsingular. 
For an arbitrary unit vector $\bm w$, we consider the linear combination $A_{\bm w}=\bm w^{\mathrm T}\bm A$, whose estimator variance is bounded by
\begin{equation}
 \operatorname{Var}(A_{\bm w})\ge \frac{1}{M} \bm w^{\mathrm T}\bm{\mathcal F}^{-1}\bm w.
\end{equation}
According to the Rayleigh quotient theorem~\cite{Rayleigh1873}, the largest value of this bound over all directions is 
\begin{equation}
 \max_{\lVert\bm w\rVert=1}\bm w^{\mathrm T}\bm{\mathcal F}^{-1}\bm w
 =\frac{1}{\lambda_{\min}(\bm{\mathcal F})}.
\end{equation}
Therefore, for fixed $M$ and probes resources, minimizing the worst-case directional variance bound is equivalent to maximizing the smallest eigenvalue of the QFIM, 
$\max_{\theta,\delta t}\lambda_{\min}[\bm{\mathcal F}(\theta,\delta t)]$. This is the E-optimality criterion~\cite{Ehrenfeld1955}, which assigns equal importance to $A_+$ and $A_\times$ on the same amplitude scale.
The optimum follows from a simple upper bound on $\lambda_{\min}$. For a real symmetric matrix,
\begin{equation}
 \lambda_{\min}(\bm{\mathcal F})
 =\min_{\lVert\bm w\rVert=1}\bm w^{\rm T}\bm{\mathcal F}\bm w,
\end{equation}
choosing $\bm w=(0,1)^{\mathsf T}$, we obtain
\begin{equation}
 \lambda_{\min}(\bm{\mathcal F})
 \le\mathcal F_{22}=\mathcal F_0\,b^2
 \le\mathcal F_0,
\end{equation}
where the last inequality follows from $b^2=\sin^2(2\theta)\sin^2\chi\le1$. For $\theta=\pi/4$ and $\chi=\pi/2$, we have $a=0$ and $b=1$, so that the QFIM reduces to
\begin{equation}
\bm{\mathcal F}=\mathcal F_0\,\mathbf I_2.
\end{equation}
Its smallest eigenvalue is therefore
\begin{equation}
\lambda_{\min}(\bm{\mathcal F})=\mathcal F_0,
\end{equation}
which saturates the upper bound and thus establishes the
E-optimality of this configuration. With $T_{\rm GW}=2\pi/\omega$, the optimal channel geometry and relative switching delay are  $\theta_{\rm opt}=\pi/4$ and $\delta t_{\rm opt}=T_{\rm GW}/4$, respectively.
\begin{figure}[t]
    \centering
    \includegraphics[width=1\linewidth, clip=true, trim=0 0 0 0]{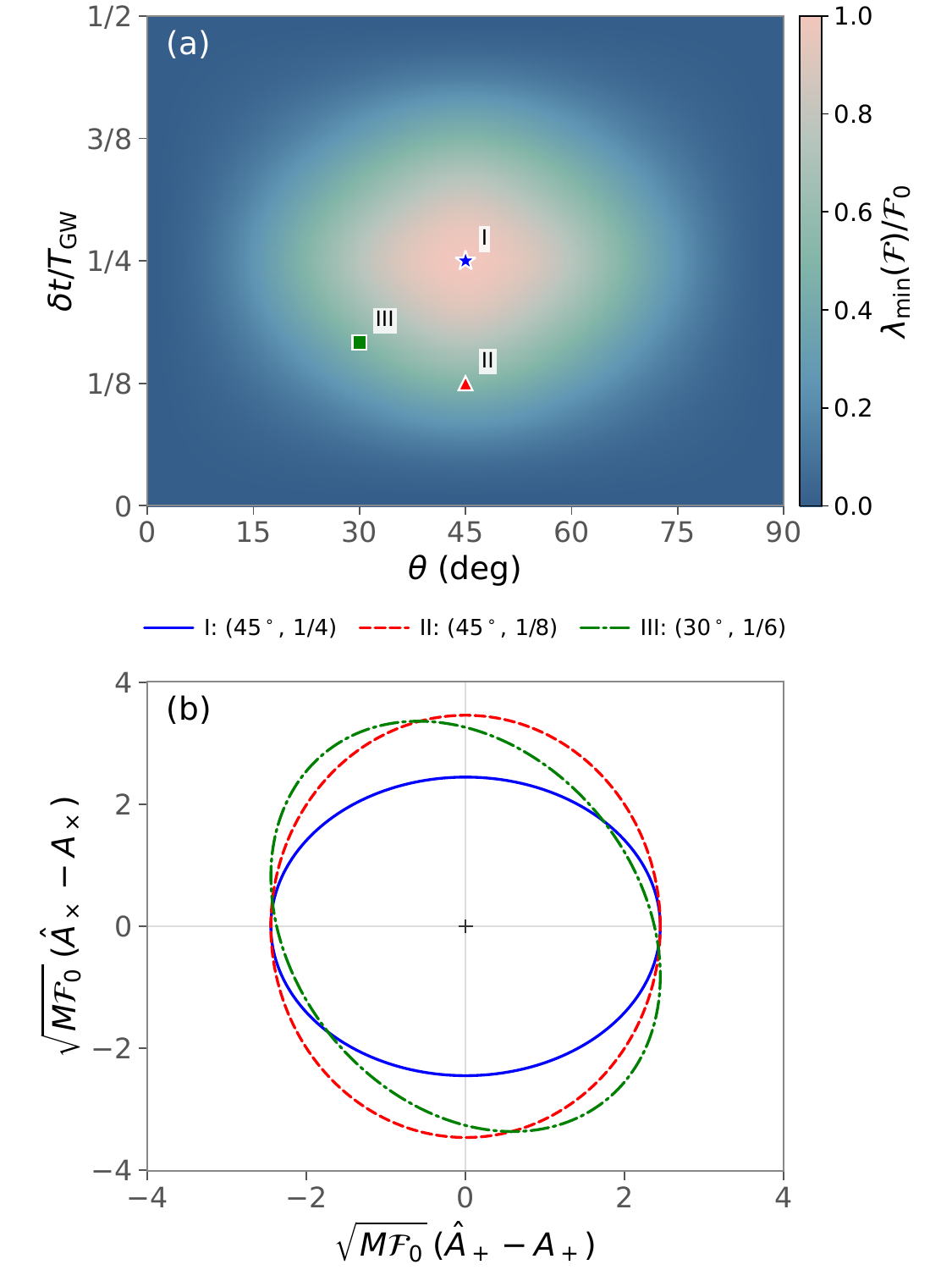}
\caption{
Joint sensitivity to the two GW polarization amplitudes as a function of channel orientation and relative switching delay. (a) Normalized smallest QFIM eigenvalue $\lambda_{\min}(\mathcal F)/\mathcal F_0$ versus $\theta$ and $\delta t/T_{\rm GW}$, with $T_{\rm GW}=2\pi/\omega$ and $\mathcal F_0=4Vg_0^2$. The marked configurations are I: $(45^\circ,1/4)$, II: $(45^\circ,1/8)$, and III: $(30^\circ,1/6)$. (b) Corresponding equal-level error ellipses associated with the matrix quantum Cramér–Rao bound, plotted in the normalized coordinates $\sqrt{M\mathcal F_0}(\hat A_+-A_+)$ and $\sqrt{M\mathcal F_0}(\hat A_\times-A_\times)$. Blue solid, red dashed, and green dash-dotted curves represent configurations I, II, and III, respectively. Configuration I gives isotropic information, $\mathcal F=\mathcal F_0 I_2$, and maximizes the information along the least-sensitive direction.
}
\label{figre51}
\end{figure}
Figure~\ref{figre51}(a) illustrates this joint optimization: configuration I attains $\lambda_{\min}/\mathcal F_0=1$, while departures from its channel angle or switching delay reduce the information along the least-sensitive direction. The corresponding contour in Fig.~\ref{figre51}(b) is isotropic, reflecting equal precision bounds for the two polarization amplitudes. In configuration II, the timing offset reduces the $\times$ response and elongates the contour along $A_\times$. In configuration III, the combined angular and timing offsets introduce an off-diagonal QFIM element and rotate the principal axes. Both II and III have $\lambda_{\min}/\mathcal F_0=1/2$, so their longest semiaxes exceed that of I by a factor of $\sqrt{2}$ at the same contour level.

The optimal configuration can be implemented using two optical channels oriented at $\gamma_1=0$ and $\gamma_2=\pi/4$, which separately probe the + and $\times$ polarization components, with $h_1(t)=h_+(t)$ and $h_2(t)=h_\times(t)$, respectively. The switching sequences are determined by the calibrated GW frequency and phase, $t_k^+ = (\phi+k\pi)/\omega$, $t_k^\times = (\phi+\pi/2+k\pi)/\omega$. The quarter-period offset synchronizes each control sequence with the corresponding polarization component.

\begin{figure}[t]
    \centering    \includegraphics[width=1\linewidth, clip=true]{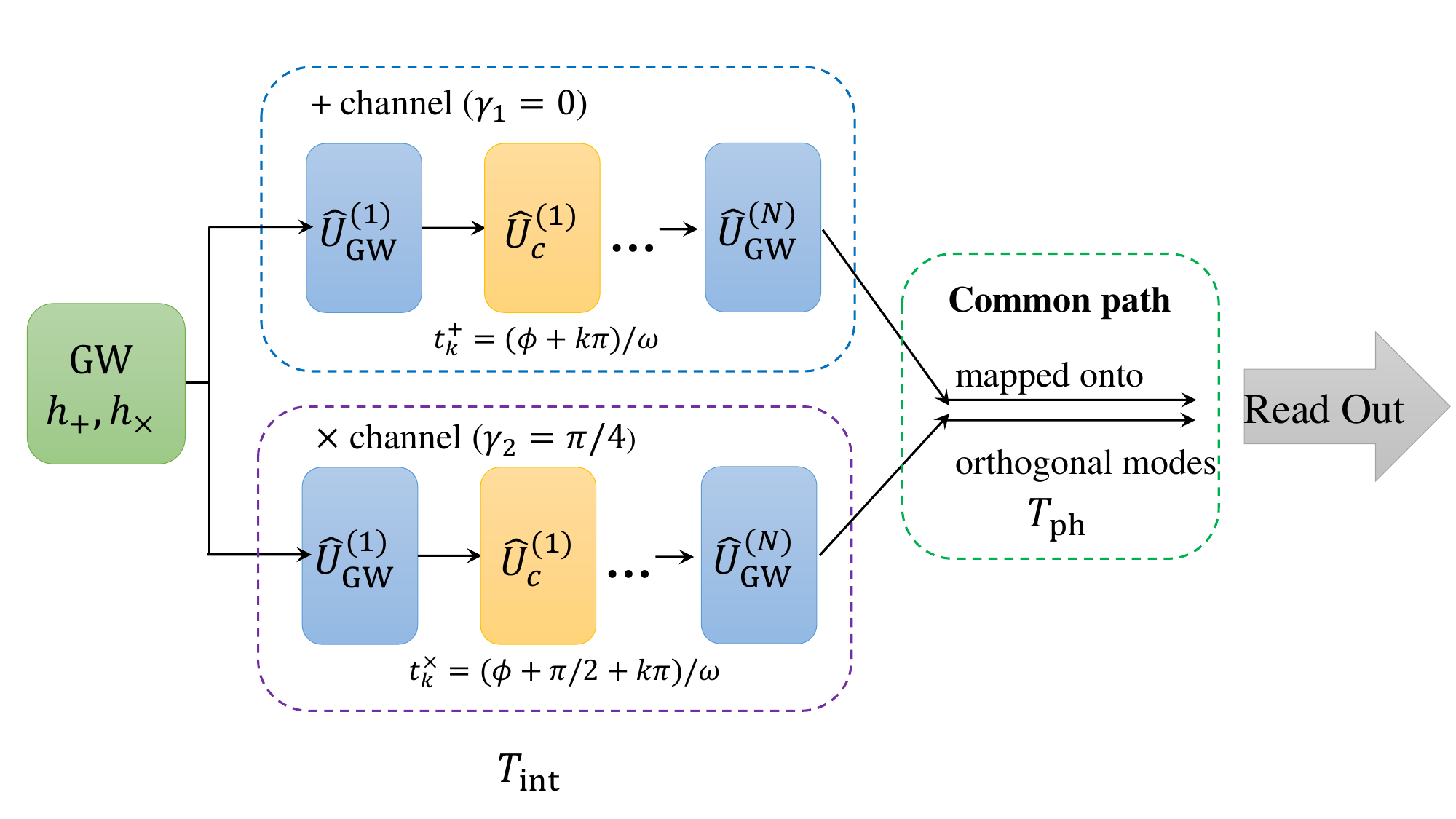}
    \caption{
    Schematic of the adaptive two-polarization GW sensing scheme. Two optical channels oriented at $\gamma_1=0$ and $\gamma_2=\pi/4$ separately probe the $+$ and $\times$ polarization components. During the interaction time $T_{\rm int}$, the GW-induced evolutions $\hat U_{\rm GW}^{(k)}$ alternate with control operations $\hat U_c^{(k)}$. The switching times are determined by the estimated GW frequency and phase, with a quarter-period offset between the two channels. The resulting frequency shifts are mapped onto orthogonal polarization modes and converted into phases $\Phi_+$ and $\Phi_\times$ during the common-path evolution time $T_{\rm ph}$. The two polarization signals are then separated and measured independently.
    }
    \label{fig3}
\end{figure}

For ideal synchronized switching, both sensing channels have the same response magnitude. 
Using the same feedback control as in the single-parameter case, the corresponding frequency shifts are $\Delta\Omega_\Theta= \Omega\mathcal R A_\Theta$.
The two GW components are sensed separately. After the GW interaction, the two signals can be mapped onto orthogonal polarization modes that propagate along the same spatial path. Over a coherent-evolution time $T_{\rm ph}$, the frequency differences are converted
into the phases $\Phi_+$ and $\Phi_\times$, which can be described by $\Phi_\Theta= \Omega\mathcal R T_{\rm ph}A_\Theta$. The sensing scheme is illustrated in Fig.~\ref{fig3}. 

At the output of the common-path delay stage, a polarization beam splitter separates two polarization modes. 
Each output can then be compared with a phase-stable local oscillator using a balanced interferometric measurement. The phases $\Phi_+$ and $\Phi_\times$ are obtained from the corresponding differential photocurrents and provide independent estimates of $A_+$ and $A_\times$.

For identical and independent probe modes, we take $\mathrm{Var}(\hat G_\Theta) = V$ and 
$\mathrm{Cov}(\hat G_+,\hat G_\times)= 0$.
Under ideal synchronization and $T_{\rm ph}\gg T_{\rm int}$, the QFIM for $\bm A=(A_+,A_\times)^\mathrm T$ then can be written as
\begin{eqnarray}
\mathbf F=4V\mathrm{diag}(g_+^2,g_\times^2) \simeq\frac{16}{\pi^2} \Omega^2\omega^2 T_{\rm int}^2T_{\rm ph}^2 V \mathcal{I},
\end{eqnarray}
where $g_\Theta=\partial\Phi_\Theta/\partial A_\Theta$, 
and $\mathcal{I}$ is the identity matrix. Both polarization amplitudes therefore have the same $T_{\rm int}^2T_{\rm ph}^2$ scaling as in the single-parameter case. The two sensing channels separate the $+$ and $\times$ responses,
while polarization multiplexing allows the two signals to share the common optical path during phase accumulation. They are separated again at the output and measured independently.

This common-path polarization scheme is experimentally attractive because most optical-path fluctuations are shared by the two modes. Its performance is mainly limited by polarization-dependent loss, birefringence, polarization crosstalk, and the coherence time of the delay or storage stage.

The ideal QFIM assumes an ideal readout of the two polarization channels. In practice, their measured phases may contain correlated noise. For  coherent- or squeezed-light readout, we use the linear model
$\bm y=\mathbf S\bm A+\bm\epsilon$. {Here
$\mathbf S=\mathrm{diag}(g_+,g_\times)$ and $\bm\epsilon$ denotes the readout noise.}

For zero-mean Gaussian noise with covariance $\mathbf\Sigma$, the classical Fisher information matrix is
\begin{equation}\label{eq:FI_two}
\mathbf I_A
=
\mathbf S^{\rm T}\mathbf\Sigma^{-1}\mathbf S .
\end{equation}
For any physical measurement, $\mathbf I_A \preceq \mathbf F $. Equality corresponds to an optimal measurement that extracts
all the information allowed by the quantum state
\cite{BraunsteinCaves1994,Holevo1982}.
The QFIM therefore provides the quantum benchmark, while
$\mathbf I_A$ determines the precision attainable with the chosen readout. The corresponding noise-weighted estimate is
\begin{equation}\label{eq:GLS}
\hat{\bm A}
=
\mathbf I_A^{-1}
\mathbf S^{\rm T}\mathbf\Sigma^{-1}\bm y,
\end{equation}
where $\mathbf I_A^{-1}=\mathrm{Cov}(\hat{\bm A})$. 
Thus, correlations between the two readout channels directly affect the precision of the joint estimation.

A common displacement contribution can be monitored by an additional reference output \cite{Gefen2024,Tarafder2025}. 
If both sensing channels contain the same displacement noise $d$ and the reference has negligible GW response, we write
\begin{subequations}
\begin{align}
y_\Theta=& g_\Theta A_\Theta+d+\epsilon_\Theta,\\
y_{\rm ref}=&d+\epsilon_{\rm ref}.
\end{align}
\end{subequations}
Subtracting the reference output gives
\begin{equation}
z_\Theta = y_\Theta-y_{\rm ref} = g_\Theta A_\Theta+\epsilon_\Theta-\epsilon_{\rm ref}.
\end{equation}
For independent measurement noises $\epsilon_+$, $\epsilon_\times$, and $\epsilon_{\rm ref}$, the covariance becomes
\begin{equation}\label{eq:ref_noise}
\mathbf\Sigma_z=
\begin{pmatrix}
\sigma_+^2+\sigma_{\rm ref}^2 &
\sigma_{\rm ref}^2\\
\sigma_{\rm ref}^2 &
\sigma_\times^2+\sigma_{\rm ref}^2
\end{pmatrix}.
\end{equation}
The reference suppresses the common displacement contribution, but also introduces additional measurement noise. Its benefit therefore depends on a small reference noise and a well-matched response to the common displacement.

The waveform estimates $(\omega,\phi)$ determine the switching times, while the measured optical phases update the estimates of both GW amplitudes. Let $\hat{\bm A}_{m-1}$ and $\mathbf P_{m-1}$ denote the current amplitude estimate and its covariance. For the $m$th measurement, the linear Gaussian updates give
\begin{equation}\label{eq:Bayes_cov}
\mathbf P_m^{-1} = \mathbf P_{m-1}^{-1} + \mathbf S_m^{\rm T} \mathbf\Sigma_m^{-1} \mathbf S_m ,
\end{equation}
and
\begin{equation}\label{eq:Bayes_mean}
\hat{\bm A}_m = \mathbf P_m \left[ \mathbf P_{m-1}^{-1}\hat{\bm A}_{m-1} + \mathbf S_m^{\rm T} \mathbf\Sigma_m^{-1}\bm y_m \right].
\end{equation}
If several calibrated readout settings are available, one can select the setting that minimizes a predicted uncertainty measure, such as
$\mathrm{Tr}(\mathbf P_m)$. The full measurement procedure is
summarized in Algorithm~\ref{alg1}.

\begin{algorithm}[t]
\caption{Adaptive two-polarization amplitude estimation}
\label{alg1}

\KwIn{
Initial estimate $(\hat{\bm A}_0,\mathbf P_0)$;
waveform estimates $(\omega,\phi)$;
maximum round $M$;
tolerance $P_{\rm tol}$
}

\KwOut{
Amplitude estimate $\hat{\bm A}$ and covariance $\mathbf P$
}

Initialize
$(\hat{\bm A},\mathbf P)
\gets
(\hat{\bm A}_0,\mathbf P_0)$\;

\For{$m=1,\ldots,M$}{

Set the switching times using
$(\omega,\phi)$\;

For each available readout setting, calculate
$\mathbf S_m$, $\mathbf\Sigma_m$, and the predicted FIM
$\mathbf I_A^{(m)}
=\mathbf S_m^{\rm T}\mathbf\Sigma_m^{-1}\mathbf S_m$\;

Choose the setting that minimizes the predicted
$\mathrm{Tr}(\mathbf P_m)$\;

Measure the two channels and record $\bm y_m$\;

Update $(\hat{\bm A},\mathbf P)$ using
Eqs.~\eqref{eq:Bayes_cov}--\eqref{eq:Bayes_mean}\;

\If{$\mathrm{Tr}(\mathbf P)<P_{\rm tol}$}{
\textbf{break}\;
}

\If{new waveform information is available}{
update $(\omega,\phi)$\;
}
}

\textbf{return} $(\hat{\bm A},\mathbf P)$\;

\end{algorithm}

For independent measurements between rounds, the above updates are exact within the linear Gaussian model. If the GW frequency and phase must be estimated jointly with the amplitudes, the problem becomes nonlinear and requires a more general Bayesian filtering procedure \cite{Tarafder2025}.

The common-path polarization layout can reduce differential optical fluctuations, but the achievable precision is still limited by unequal losses, polarization mixing, mirror motion, reference noise, and the finite coherence time of the phase-accumulation stage. For twin-Fock probes, a suitable photon-number-resolving measurement is required to assess whether the ideal QFIM can be attained.

\section{Conclusion}\label{section 5}

We have developed an adaptive quantum-metrology framework for probing weak gravitational waves via their direct energy exchange with light. 
The central idea is to use feedback-controlled SU(2) operations to synchronize the optical propagation-direction switching with the estimated GW waveform. This prevents the frequency shifts acquired in successive parts of the GW cycle from cancelling and allows the GW response to accumulate constructively over the interaction time. 

The scheme separates the sensing process into two stages. During the interaction time $T_{\rm int}$, the GW produces a persistent optical-frequency difference, with the accumulated response scaling as $\mathcal R \propto \omega T_{\rm int}$  in the ideal case. The remaining frequency difference is converted into a relative phase during a coherent phase-accumulation time $T_{\rm ph}$. The corresponding QFI scales as $F_A \propto T_{\rm int}^2T_{\rm ph}^2 \mathrm{Var}(\hat G)$. The adaptive control preserves the time-dependent response, while the probe state determines the quantum enhancement through
$\mathrm{Var}(\hat G)$. For a fixed-photon-number nonclassical probe, this gives a Heisenberg-like $1/n$ scaling.

We have further generalized the framework to the simultaneous estimation of the two GW polarization amplitudes. The joint sensitivity depends on both the relative orientation and the switching delay of the two sensing channels. Optimizing the smallest eigenvalue of the QFIM gives two channels separated by $45^\circ$ and switching sequences shifted by one quarter of a GW period.In this configuration, the QFIM becomes diagonal and gives equal precision for $A_+$ and $A_\times$. The two accumulated frequency shifts can then be mapped onto orthogonal polarization modes and converted into phases along a common optical path. This provides a simple optical implementation of polarization-resolved multiparameter sensing.

The QFIM sets the quantum precision bound, while the classical Fisher information characterizes the precision attainable with a specified readout. This framework also accomondates correlated readout noise and reference-assisted measurements. The adaptive estimation procedure can be interpreted as a sequential accumulation of Fisher information, with updated waveform information used to preserve a large response in subsequent measurements.

The present framework combines GW--light energy exchange, adaptive control, post-interaction phase accumulation, and polarization-resolved multiparameter estimation within a single metrological description. Future work may include optical loss, imperfect switching, and polarization-dependent noise, as well as   optimization of the common-path readout. Extending the parameter set to include the GW frequency and phase would connect the present scheme toward adaptive quantum waveform estimation.

\section{Acknowledgments}
We would like to thank Prof. Qing Ai for helpful discussions. This work was supported by the Fundamental and Interdisciplinary Disciplines Breakthrough Plan of the Ministry of Education of China under Grant ( No. JYB2025XDXM115, JYB2025XDXM201), Guangdong Provincial Quantum Science Strategic Initiative under Grant No.~GDZX2505004, National Natural Science Foundation of China under Grant No.~62461160263, and Quantum Science and Technology-National Science and Technology Major Project (2023ZD0300200).
\appendix
\nocite{Graefe2023}
\bibliography{GW_adaptive_refs}

\end{document}